\documentclass[%
 reprint,
 amsmath,amssymb,
 aps,
]{revtex4-2}

\usepackage{graphicx}
\usepackage{dcolumn}
\usepackage{lipsum}
\usepackage{bm}
\usepackage{xcolor}

\usepackage[normalem]{ulem}

\begin{document}

\preprint{APS/123-QED}

\title{Density--Velocity Relation Is Scale-Dependent in Epithelial Monolayers}

\author{Hengdong Lu}
 \author{Tianxiang Ma}
\author{Amin Doostmohammadi}%
 \email{doostmohammadi@nbi.ku.dk}
\affiliation{Niels Bohr Institute, University of Copenhagen, Blegdamsvej 17, Copenhagen 2100, Denmark
}%

\date{\today}

\begin{abstract}
The relationship between cell density and velocity is often assumed to be negative, reflecting crowding-induced suppression of movement. However, observations across systems reveal a more nuanced picture: while some emphasize contact inhibition of locomotion, others suggest that dense regions exhibit enhanced activity due to force generation and stress buildup.
Here, using experimental measurements we show that density--velocity relations in epithelial monolayers are inherently scale dependent.
By coarse-graining cell trajectories over multiple spatial windows, we find that cell velocity correlates positively with local density at small scales, but negatively at large scales. Employing traction force measurements, we find that this crossover coincides with the emergence of mechanical pressure segregation, defining a characteristic length scale beyond which crowding dominates. A minimal model incorporating activity-induced shape changes reproduces this crossover and identifies the competition between active force generation and mechanical confinement as the underlying mechanism. Our results reconcile conflicting views of density-regulated migration and highlight an emergent length scale as a key factor in interpreting collective cell dynamics.
\end{abstract}

\maketitle

\section{Introduction}
In collective cell systems, cell density and velocity are fundamental quantities that shape how tissues grow, migrate, and reorganize \cite{atiaAreCellJamming2021a, lawson2021jamming,cheung2025collective,lin2025gradients}. A common assumption is that these two quantities are negatively correlated: as density increases, crowding and cell--cell contact suppress individual motion. This view, embodied in the concept of contact inhibition of locomotion \cite{abercrombie1970contact,theveneau2010collective,puliafito2012collective,zimmermann2016contact}, has provided a useful framework for explaining density-regulated processes such as wound healing, tissue spreading, and cancer invasion \cite{li2013collective,tian2024motility}. 

At the same time, 
a growing number of experimental studies reported the different trend, where dense regions exhibit enhanced activity and more higher local velocity~\cite{parkUnjammingCellShape2015a,xuSelfenhancedMobilityEnables2024,chisolmTransitionsCooperativeCrowdingdominated2025,ruider2024topological}. Consistent with this, measurements of traction forces and stresses suggest that increased density coincides with stronger mechanical activity \cite{serra2012mechanical, shahin2020tissue}, and mechanosensitive pathways reinforce contractility under compression \cite{loza2016cell}. 

On the theoretical side, continuum models of active matter often assume that active stresses scale directly with cell density, predicting stronger collective flows in denser regions \cite{thampi2015intrinsic,doostmohammadi2015celebrating}. Conversely, particle-based models emphasize a negative feedback between density and speed, which underlies motility-induced phase separation (MIPS) in synthetic active systems \cite{fuStripeFormationBacterial2012,cates2015motility}. 
As such, the density--velocity relation is usually treated as monotonic, either uniformly negative or uniformly positive
\cite{catesArrestedPhaseSeparation2010,marchetti2013hydrodynamics,roycroftMolecularBasisContact2016}. The apparent contradiction between negative correlations emphasized in contact inhibition and positive correlations reported in recent studies points to a missing principle. Resolving this discrepancy requires moving beyond a uniform view of density–velocity coupling.

We propose that the missing ingredient is the spatial scale. While local interactions can promote motion through active force generation and shape changes \cite{PhysRevLett.122.048004}, global crowding can suppress motion through mechanical confinement
\cite{angelini2011glass,atia2018geometric,lawson2021jamming}. If so, the sign of the density--velocity relations should depend on the spatial window of observation. Yet systematic tests of this possibility have been lacking, in part because of the challenge of disentangling small-scale fluctuations from large-scale gradients in cell populations. We address this issue by quantifying density--velocity relations across spatial scales in epithelial monolayers. By systematically varying the size of local windows, we computed both the locally averaged speed and the corresponding local density. This analysis revealed a robust crossover: cellular velocity correlates positively with density at small scales but negatively at large scales. Using Bayesian Inversion Stress Microscopy (BISM), we further measured the stress distribution within the cell collectives and found that the transition length scale of density--velocity relations coincides with the characteristic scale of pressure segregation, where mechanical heterogeneity emerges. Minimal modeling of cell collectives incorporating activity-induced shape deformations recapitulates this crossover and identifies the competition between active force generation and mechanical confinement as the underlying mechanism. Together, these results establish scale as a key organizing principle for density--velocity coupling and reconcile conflicting interpretations of collective cell migration.

\section{Methods}

\noindent{\bf Experimental setup.}
Madin–Darby Canine Kidney (MDCK) epithelial cells were cultured in Dulbecco’s Modified Eagle Medium (DMEM) supplemented with 10\% fetal bovine serum and 1\% penicillin–streptomycin under standard conditions. Cells were seeded onto Nunclon\textsuperscript{TM} Delta-treated plastic dishes (Thermo Scientific\texttrademark) and allowed to reach confluence within 24--36~h.  

Time-lapse phase-contrast imaging was performed on a Nikon ECLIPSE Ti microscope equipped with a temperature- and CO\textsubscript{2}-controlled incubation chamber (H201-K-FRAME, Okolab), maintaining cells at 37\,\textdegree C and 5\% CO\textsubscript{2}. Images were acquired at 10$\times$ magnification every 10~min for a total of 8--10~h. Velocity fields were computed using Particle Image Velocimetry (PIV) with interrogation windows of 64\,$\times$\,64 and 32\,$\times$\,32 pixels and 50\% overlap, following Thielicke and Stamhuis \cite{thielickePIVlabUserfriendlyAffordable2014}.  

Local cell density was quantified using the CellSegmentationTracker Python module (\url{https://github.com/simonguld/CellSegmentationTracker}), which segments individual cells and computes density by counting cells inside a circular window of adjustable diameter $d$ centered at each spatial location.  

To measure intercellular isotropic pressure, we employed Bayesian Inversion Stress Microscopy (BISM) \cite{nier2016inference}. Substrates were embedded with fluorescent microspheres, which were imaged by epi-fluorescence during cell culture. A reference image of the relaxed substrate was acquired at the end of each experiment by adding 200~µL of 10\% sodium dodecyl sulfate (SDS) or 10\% Triton X-100 to detach cells. Bead images were then registered using the Image Stabilizer plugin in FIJI \cite{li2008image}, and illumination correction was applied to reduce background noise. Displacement fields were extracted by pairwise PIV analysis in PIVlab using 32\,$\times$\,32 pixel interrogation windows with 50\% overlap.  

Traction force fields were calculated via Fourier Transform Traction Cytometry (FTTC), and stress tensor fields were inferred using the BISM algorithm \cite{NIER20161625}. The isotropic intercellular pressure was computed as half the trace of the stress tensor, $(\sigma_{xx} + \sigma_{yy})/2$, with positive values denoting tension and negative values compression. Further methodological details on BISM are provided in the Supplementary Information.  

\noindent{\bf Multi-phase field model of epithelial monolayers.}  
We employed a multi-phase field model of epithelial monolayers to simulate the collective behaviors of MDCK cell layers \cite{PhysRevLett.122.048004,muellerPhaseFieldModels2021,monfared2025multi,PhysRevE.110.044403}. Briefly, each cell is represented by a phase field $\phi_i$, evolving according to
\begin{equation}
    \partial_t \phi_i + \mathbf{v}_i \cdot \nabla \phi_i = - \frac{\delta \mathcal{F}}{\delta \phi_i}, \qquad i=1,\dots,N ,
    \label{eq:dynamics}
\end{equation}
where $\mathcal{F}$ is the free-energy functional and $\mathbf{v}_i$ is the velocity of cell $i$.  

The free energy combines passive contributions,  
\begin{equation}
\mathcal{F} = \mathcal{F}_{\text{CH}} + \mathcal{F}_{\text{area}} + \mathcal{F}_{\text{rep}} + \mathcal{F}_{\text{adh}},
\end{equation}
with terms defined as
\begin{equation}
\begin{aligned}
\mathcal{F}_{\text{CH}} &= \sum_{i} \frac{\gamma}{\lambda} \int d\mathbf{r} \left[ 4\phi_i^2(1-\phi_i)^2 + \lambda^2 (\nabla \phi_i)^2 \right], \\
\mathcal{F}_{\text{area}} &= \sum_{i} \mu \left( 1 - \frac{1}{\pi R^2} \int d\mathbf{r}\, \phi_i \right)^2, \\
\mathcal{F}_{\text{rep}} &= \sum_{i} \sum_{j \neq i} \frac{\kappa}{\lambda} \int d\mathbf{r}\, \phi_i^2 \phi_j^2, \\
\mathcal{F}_{\text{adh}} &= \sum_{i}\sum_{j \neq i} \omega\lambda\int d\mathbf{r}\,\nabla\phi_i \cdot \nabla\phi_j .
\end{aligned}
\label{eq:freeenergy}
\end{equation}

\noindent Here, $\mathcal{F}_{\text{CH}}$ is Cahn-Hilliard free energy which stabilizes cell interfaces
\cite{palmieriMultipleScaleModel2015}. $\mathcal{F}_{\text{area}}$ enforces a soft constraint on cell area to be close to the initially assigned circular one, $\pi R^2$. Still, cell area is possibly changed by squeezing by neighbors
\cite{peyretSustainedOscillationsEpithelial2019}. $\mathcal{F}_{\text{rep}}$ penalizes cell overlap, and $\mathcal{F}_{\text{adh}}$ promotes adhesion. The interface width is set by $\lambda$, and the coefficients $\gamma, \mu, \kappa, \omega$ control the strength of shape stabilization, area elasticity, repulsion, and adhesion, respectively.  

At the cellular level, intercellular forces drive cell motion in an overdamped regime,
\begin{equation}
    \xi \mathbf{v}_i = \mathbf{F}_i^{\text{int}},
\end{equation}
where $\xi$ is the substrate friction coefficient and $\mathbf{F}_i^{\text{int}}$ is the net intercellular force. This force is obtained from the tissue stress tensor $\sigma_{\text{tissue}}$,  
\begin{equation}
    \mathbf{F}_i^{\text{int}} = \int d\mathbf{x}\, \phi_i \nabla \cdot \sigma_{\text{tissue}} = -\int d\mathbf{x}\, \sigma_{\text{tissue}} \cdot \nabla \phi_i ,
\end{equation}
with
\begin{equation}
    \sigma_{\text{tissue}} = \sigma_{\text{passive}} + \sigma_{\text{active}} .
\end{equation}

The passive stress derives from the free energy \cite{Cates_Tjhung_2018, PhysRevE.110.044403},
\begin{align}
    f_{\text{passive}} &= \nabla \cdot \sigma_{\text{passive}} = \frac{\delta \mathcal{F}}{\delta \phi_i}\nabla \phi_i , \\
    \sigma_{\text{passive}} &= -P \mathbb{I} - \sum_i \nabla \phi_i \frac{\partial \mathcal{F}}{\partial \nabla \phi_i}, \\
    P &= -\sum_i f_i - \mu_i \phi_i ,
\end{align}
where $f_{\text{passive}}$ is the passive force density, $f_i$ is the free-energy density of $\phi_i$, $\mu_i = \delta \mathcal{F}/\delta \phi_i$ denotes the chemical potential, and $\mathbb{I}$ is the identity tensor. The isotropic pressure is given by the scalar contribution of the passive stress.  

For active processes, we adopt a minimal model that includes extensile shape deformations,
\begin{equation}
    \sigma_{\text{active}} = -\zeta \sum_i \phi_i \mathbf{S}_i ,
\end{equation}
with activity strength $\zeta > 0$ and shape tensor
\begin{equation}
    \mathbf{S}_i = -\int d\mathbf{x} \left[ \nabla \phi_i \nabla \phi_i^T - \tfrac{1}{2}\text{Tr}(\nabla \phi_i \nabla \phi_i^T)\right].
\end{equation}
This term captures the active interaction in which a cell pushes neighbors along its elongation axis and pulls them along its contraction axis.  

The parameters used in this study are consistent with physiological ranges \cite{monfared2025multi} and are listed in \textbf{Appendix Table~1}.

\begin{figure}[hb!]
    \centering
    \includegraphics[width=1\linewidth]{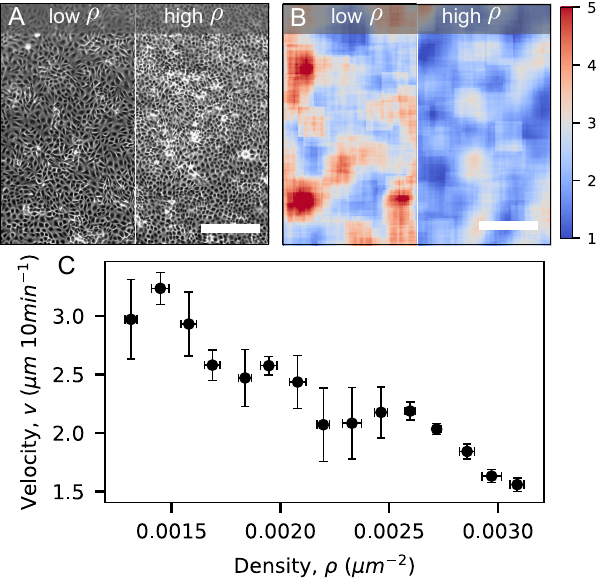}
    \caption{\textbf{Increased global cell density suppresses cell motion.} 
    \textbf{(A)} Representative bright-field images of MDCK monolayers at low (left) and high (right) global cell density ($\rho$). Scale bar: 200 $\mu m$.
    \textbf{(B)} Corresponding velocity magnitude maps reveal that low-density cell layers exhibit higher speed, while high-density cell layers show suppressed motion. Scale bar: 200 $\mu m$. Unit of colar bar: $\mu m/ 10\mathrm{mins}$
    \textbf{(C)} Quantification of average cell velocity as a function of global cell density ($\rho$), showing a negative correlation (mean ± SD, $n = 4$ independent experiments).
    }
    \label{fig:1}
\end{figure}

\begin{figure*}
    \centering
    \includegraphics[width=1\linewidth]{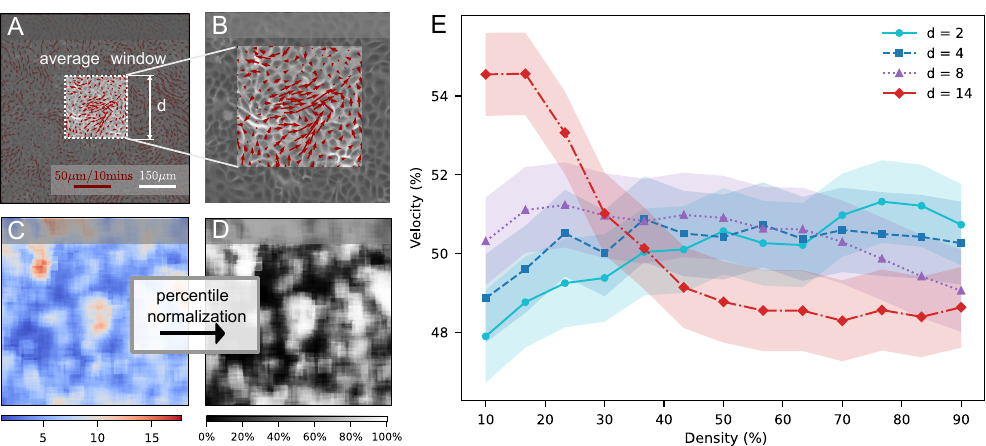}
    \caption{\textbf{Scale-dependent local relationships between cell velocity and density.} Density and velocity magnitude field are coarse-grained calculated as the average over cells within the square window with size $d\times d$ \textbf{(A, B)}. The original field is percentile normalized according to the instantaneous ranking among all positions. This normalization eliminates two sources of fluctuation: the global contribution, and the relative fluctuations between local regions \textbf{(C, D)}. Tuning the average window size d, we get density-velocity correlation at multiple scales. The length scale $d$ is in cell size units. Density-velocity relations is scale-dependent. As d increases, correlation between density and velocity changes from positive to negative \textbf{(E)}. The unit of d is cell diameter (mean ± SEM, $n = 4$ independent experiments).}
    \label{fig:2}
\end{figure*}

\section{Results}

\noindent{\bf Global density suppresses cell motion.}  
Previous studies have highlighted the role of jamming in epithelial monolayers, showing that as tissues approach the jamming transition, crowding suppresses motion and leads to a dynamical slowdown at high density \cite{angelini2011glass,garciaPhysicsActiveJamming2015,lawson2021jamming}. In our confluent system, we therefore expected a global negative correlation between cell density and velocity.  

To test this, we defined an observation window of $800~\mu$m $\times$ $800~\mu$m, corresponding to approximately 33 cell diameters. Global cell density $\rho$ was measured by counting the number of cells within the window and dividing by the area, while global velocity was calculated as the mean speed of cells in the same region. We analyzed four independent experimental datasets, each containing 70 consecutive frames acquired during continuous monolayer growth, providing a broad range of densities as the system evolved. To reduce noise, we combined frames from all datasets and grouped them into bins based on their global density $\rho$. For each bin, we calculated the mean velocity $v$ across all frames falling within the corresponding density range. Error bars in Fig.~\ref{fig:1}C indicate the variance of both $\rho$ and $v$ within each bin.

Consistent with predictions from jamming dynamics\cite{angelini2011glass,garciaPhysicsActiveJamming2015,lawson2021jamming}, we observed a robust negative correlation between global density and velocity (Fig.~\ref{fig:1}C). This trend was also evident qualitatively: representative snapshots and velocity magnitude maps (Fig.~\ref{fig:1}A,B) show that monolayers with higher global density consistently exhibited slower motion than less dense ones, even though both density and velocity fields remained highly heterogeneous at the local scale.\\

\noindent{\bf Density--velocity relation is scale-dependent.} The global analysis suggests that high density suppresses motion, but such negative correlations are obtained only when averaging over large windows spanning tens of cell diameters. Given the strong spatial heterogeneity within epithelial monolayers \cite{zehnderMulticellularDensityFluctuations2015,vishwakarmaWhyDoesEpithelia2019}, it is unclear whether this relation holds at smaller scales.  

To probe the instantaneous local coupling between density and velocity, we constructed coarse-grained spatial fields of both quantities. At each location, local density was defined as the number of cells within a surrounding $d \times d$ window, and local velocity as the average cell speed in the same region. To isolate genuine local effects, we applied a percentile normalization procedure to remove correlations driven by global density fluctuations (Fig.~\ref{fig:2}C, D). By systematically varying the window size $d$, we quantified the density–velocity relations across scales.  

This analysis revealed a striking scale dependence (Fig.~\ref{fig:2}E). At large window sizes (above $\sim$14 cell diameters), the global negative correlation between density and velocity was recovered. In contrast, at small scales, we observed the opposite trend: local density and motion were weakly but consistently positively correlated. As the observation window increased, a gradual crossover from positive to negative correlation emerged.  

Together, these results establish a multi-scale picture: at large scales, crowding suppresses motion, whereas at small scales, dense regions tend to exhibit enhanced dynamics relative to their surroundings.\\

\noindent{\bf Pressure segregation correlates with the crossover length scale.}  
The emergence of opposite correlations at different scales raises the question of what mechanism underlies this crossover. In particular, why do dense regions locally support faster dynamics, yet globally exhibit suppressed motion? To address this, we next examined the spatial organization of mechanical pressure within the monolayer. Because pressure reflects how local force generation and global confinement balance each other
\cite{tambeCollectiveCellGuidance2011}, its segregation provides a natural candidate mechanism for the observed scale dependence of density–velocity correlations.

\begin{figure}[t!]
    \centering
    \includegraphics[width=1\linewidth]{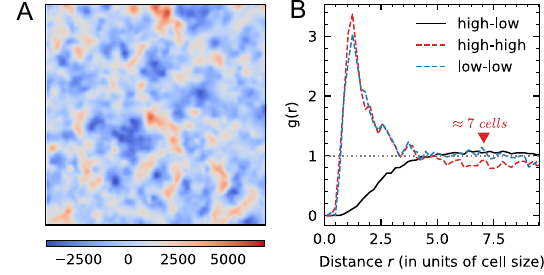}
    \caption{\textbf{Mechanical pressure in MDCK tissue is heterogeneous.} High pressure regions are spatially segregated from low pressure regions \textbf{(A)}. The spacing between high- and low- pressure regions gives the length scale of pressure heterogeneity. Cross-type radial distribution function $g(r)$ verify the depletion between cell with high and low pressure. The spacing length scale is given 7 cells as point cross $g(r)$ reaches maximum \textbf{(B)}. Averaged across $n = 4$ independent experiments. 
    }
    \label{fig3}
\end{figure}

We quantified isotropic pressure in confluent MDCK monolayers using Bayesian Inversion Stress Microscopy (BISM) \cite{nier2016inference}. The resulting pressure maps revealed pronounced heterogeneity, characterized by spatially correlated high- and low-pressure regions (Fig.~\ref{fig3}A). This suggested the presence of a characteristic length scale, defined by the typical spacing between segregated regions.  

To determine this length scale, we computed the cross-type radial distribution function $g(r)$ between high-pressure and low-pressure cells
\cite{brankaPairCorrelationFunction2011}. Each cell was assigned a pressure value based on the mean isotropic pressure within one cell radius ($\sim$12~µm) around its center of mass. Cells in the top 25\% of the pressure distribution were labeled high-pressure, while those in the bottom 25\% were labeled low-pressure. The precise choice of threshold did not affect the analysis (see \textbf{Supplementary Fig. S1}). The resulting pair distributions $g_{\text{high--high}}(r)$ and $g_{\text{low--low}}(r)$ showed features typical of disordered liquid and colloidal systems: a pronounced nearest-neighbor peak at one cell diameter followed by oscillatory decay towards unity
\cite{brankaPairCorrelationFunction2011}. In contrast, the cross distribution $g_{\text{cross}}(r)$ displayed a clear depletion zone extending over several cell diameters, indicating spatial segregation of high- and low-pressure regions. We defined the characteristic segregation length as the distance at which $g_{\text{cross}}(r)$ first rose above unity, yielding a value of approximately seven cell diameters (Fig.~\ref{fig3}B). Notably, this length scale coincides with the crossover scale at which the local density–velocity relation changes sign from positive to negative (Fig.~\ref{fig:2}E), indicating that mechanical pressure segregation correlates with the transition between local enhancement and global suppression of cell motion.


\noindent{\bf Minimal cell-based model reveals mechanism of scale-dependent coupling and pressure segregation.}  
To test whether the scale-dependent density–velocity relation and the associated emergence of pressure segregation arise from general principles, we constructed a minimal phase-field model of a confluent monolayer. In this model, the only source of activity is extensile shape deformation, which allows cells to exert forces along their elongation axis
\cite{PhysRevLett.122.048004,muellerPhaseFieldModels2021,PhysRevE.110.044403}. Despite its simplicity, the model reproduces the essential features of experiments~\cite{monfared2025multi}, enabling a mechanistic interpretation of the observed behavior.  

\begin{figure}
    \centering
    \includegraphics[width=1\linewidth]{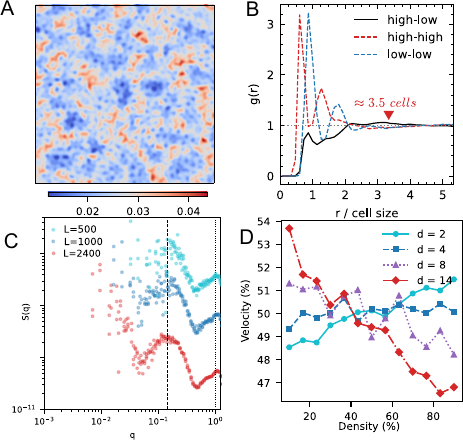}
    \caption{\textbf{Simulations reproduce pressure segregation and its link to the density–velocity crossover.} In simulation systems, pressure segregation is also observed \textbf{(A)}. Depletion in cross \(g(r)\) giving a length scale $\approx3.5\ cells$ \textbf{(B)}. Structure factor $S(q)$ of mechanical pressure is calculated. For super large-scale simulation of 40000 cells, low-frequency peak in $S(q)$ gives the same length scale \textbf{(C)}. Density-velocity relation turns from positive to negative at several cells size, close to length scale of mechanical pressure \textbf{(D)}. When window size is greater that length scale of mechanical pressure, cell crowding dominates and suppresses cell motion at large scale. Averaged across $n = 3$ independent simulations.}
    \label{fig:4}
\end{figure}

Simulations revealed a scale-dependent density–velocity correlation closely matching experimental data (Fig.~\ref{fig:4}D). At small observation windows of a few cell diameters, regions of higher density exhibited enhanced velocity, whereas at larger windows, velocity was suppressed in dense regions. The crossover occurred at a window size of $\sim$3.5 cell diameters. This demonstrates that activity-induced shape deformations alone are sufficient to generate the transition from local enhancement to global suppression.  

A key advantage of the model is that isotropic pressure can be computed directly from the stress tensor
\cite{muellerPhaseFieldModels2021,PhysRevE.110.044403}. Simulated pressure maps exhibited spontaneous heterogeneity, with domains of high and low pressure emerging dynamically (Fig.~\ref{fig:4}A). Quantitative analysis using the cross-type radial distribution function $g(r)$ confirmed the segregation of pressure, revealing a depletion zone with a characteristic length scale of $\sim$3.5 cell diameters (Fig.~\ref{fig:4}B), in good agreement with the crossover scale of the density–velocity relation.  

To further validate this emergent length scale, we computed the structure factor $S(q)$ of the isotropic pressure field,
\begin{equation}
    S(\mathbf{q}) = \int e^{-i\mathbf{q}\cdot\mathbf{r}} C(\mathbf{r})\, d\mathbf{r},
\end{equation}
where $C(\mathbf{r})$ is the spatial autocorrelation function, and $S(q)$ is the angular average of $S(\mathbf{q})$. To overcome noise at low frequencies, we performed simulations of increasingly large systems, reaching up to 40,000 cells. As shown in Fig.~\ref{fig:4}C, the resulting $S(q)$ displayed two characteristic peaks. At high frequencies, a peak corresponding to one cell diameter reflects the intrinsic single-cell scale. At lower frequencies, a second peak appeared at a wavelength of $\sim$7 cell diameters—twice the segregation length identified from $g_{\text{cross}}(r)$. This correspondence indicates that pressure segregation organizes into alternating high- and low-pressure domains, whose periodicity is directly captured by the structure factor. These findings suggest a clear mechanistic picture that can be made explicit by examining how activity and confinement compete across scales. We therefore turn to a mechanistic analysis.

\section{Mechanism}
In principle, the relationship between cell speed and density can be formulated as a functional \cite{cates2015motility},
\begin{equation}
    v = v([\rho]),
\end{equation}
where $[\rho]$ denotes the full spatial field of cell density. Most theoretical approaches simplify this to a purely local dependence \cite{catesWhenAreActive2013},
\begin{equation}
    v = v(\rho).
\end{equation}
A direct implication of this approximation is that coarse-graining over windows of different sizes should rescale the amplitude of correlations but not alter their sign. Our measurements directly contradict this prediction. We find that at small windows, denser regions exhibit faster motion, whereas at large windows, density suppresses velocity. This crossover demonstrates that the local approximation fails, and that the density--velocity relation is inherently multiscale.

Our minimal model provides a mechanistic explanation for this scale dependence. When local density increases, cells form more contacts, and in an extensile active system this enhances active force generation. At small scales, this results in a positive density–velocity relation. At larger scales, however, confluent tissues are mechanically confined by surrounding neighbors. Passive forces dissipate activity, and high-density domains become mechanically restricted regions where cell speed is suppressed \cite{angelini2011glass}. The observed scale dependence therefore emerges from a competition between local activity-driven force generation and global mechanical confinement.

The crossover scale at which the correlation changes sign reflects the effective length scale of mechanical confinement. Independent measurements of isotropic pressure maps confirm this interpretation: pressure segregates into high- and low-pressure domains whose spacing defines a characteristic length. This pressure-segregation length coincides with the crossover length obtained from density–velocity correlations, and is also captured in reciprocal space by the low-frequency peak of the pressure structure factor. Together, these results identify the confinement length as a unifying scale that controls the transition from local enhancement to global suppression.

\begin{figure}
    \centering
    \includegraphics[width=1\linewidth]{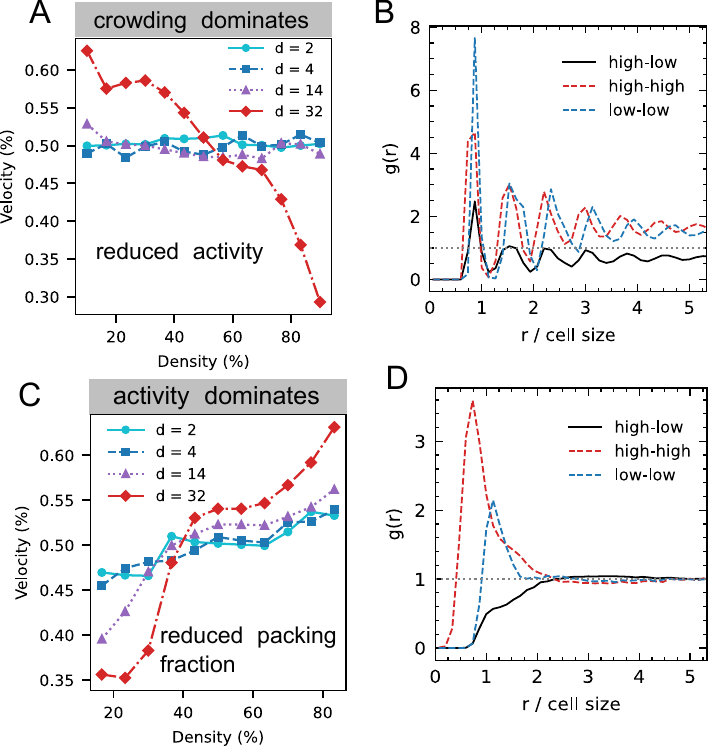}
    \caption{\textbf{Competition between active force generation and mechanical confinement leads to the scale-dependent density-velocity correlation.} By selectively weakening either activity or intercellular crowding, we access dynamical regimes in which a single mechanism dominates, validating this competition picture. \textbf{(A)} Reducing activity \(\zeta\) eliminates the small-scale positive correlation. At low activity, \(g(r)\) exhibits multiple peaks, characterizing a system close to ordered lattice structure \textbf{(B)}. The pressure segregation length diverges. The suppression of velocity by density only occurs at super large-scale (\(\sim\) 32 cell diameters). \textbf{(C)} Lowering the packing fraction weakens cell crowding. The large-scale suppression disappears and the correlation becomes positive across all scales. In this regime, high-high and low-low \(g(r)\) exhibit a standard disordered style, and the cross \(g(r)\) reveals a finite pressure segregation length scale \textbf{(D)}.}
    \label{fig:5}
\end{figure}

This two-scale mechanism is further validated by perturbing either side of the competition. Reducing the activity parameter $\zeta$ weakens active force generation. In this regime, the small-scale positive correlation disappears, and density suppresses velocity only at very large windows (Fig.~\ref{fig:5}A). The enlarged crossover scale arises because, at low activity, the tissue approaches a lattice-like ordered state, and the pressure segregation length diverges (Fig.~\ref{fig:5}B). Conversely, weakening crowding by lowering the packing fraction diminishes mechanical confinement. At reduced packing, the monolayer remains confluent and retains sufficient contact for activity, but the large-scale negative correlation vanishes, leaving a positive density–velocity relation across all accessible scales (Fig.~\ref{fig:5}C). In this regime, $g_{\text{high--high}}(r)$ and $g_{\text{low--low}}(r)$ show standard disordered features, while $g_{\text{cross}}(r)$ still reveals a finite segregation length (Fig.~\ref{fig:5}D).  

Altogether, these perturbations support the mechanistic picture that scale-dependent density–velocity coupling arises from the interplay between local activity and large-scale mechanical confinement, with the crossover scale set by the pressure-segregation length.
\section{Conclusion}

Our results establish spatial scale as a missing principle that reconciles conflicting reports of density–velocity relations in epithelial tissues. We identify an emergent length scale, set by the segregation of mechanical pressure, that determines the crossover from local enhancement to global suppression of motion. This length scale provides a unifying framework for understanding why dense regions can be simultaneously active at the microscale and arrested at the tissue scale.  

Beyond reconciling existing observations, our findings suggest new directions. The segregation length may vary across cell types, matrix environments, or pathological states, offering a physical parameter to compare different tissues. It also provides a measurable target for probing how biochemical regulation tunes mechanical heterogeneity. More broadly, the finding that density–velocity coupling is inherently scale-dependent underscores the need for multiscale approaches in active matter and tissue mechanics. Future studies should test whether similar crossovers organize processes such as morphogenesis, wound healing, and cancer invasion, where density-regulated migration plays a central role.

\begin{acknowledgments}
It is a pleasure to acknowledge helpful conversations with Andreas Bausch, Iris Ruider, and Kristian Thijssen. A. D. acknowledges funding from the Novo Nordisk Foundation (grant No. NNF18SA0035142 and NERD grant No. NNF21OC0068687), Villum Fonden (Grant no. 29476), and the European Union (ERC, PhysCoMeT, 101041418). Views and opinions expressed are however those of the authors only and do not necessarily reflect those of the European Union or the European Research Council. Neither the European Union nor the granting authority can be held responsible for them.
\end{acknowledgments}




\bibliography{apssamp}

\begin{thebibliography}{43}%
\makeatletter
\providecommand \@ifxundefined [1]{%
 \@ifx{#1\undefined}
}%
\providecommand \@ifnum [1]{%
 \ifnum #1\expandafter \@firstoftwo
 \else \expandafter \@secondoftwo
 \fi
}%
\providecommand \@ifx [1]{%
 \ifx #1\expandafter \@firstoftwo
 \else \expandafter \@secondoftwo
 \fi
}%
\providecommand \natexlab [1]{#1}%
\providecommand \enquote  [1]{``#1''}%
\providecommand \bibnamefont  [1]{#1}%
\providecommand \bibfnamefont [1]{#1}%
\providecommand \citenamefont [1]{#1}%
\providecommand \href@noop [0]{\@secondoftwo}%
\providecommand \href [0]{\begingroup \@sanitize@url \@href}%
\providecommand \@href[1]{\@@startlink{#1}\@@href}%
\providecommand \@@href[1]{\endgroup#1\@@endlink}%
\providecommand \@sanitize@url [0]{\catcode `\\12\catcode `\$12\catcode `\&12\catcode `\#12\catcode `\^12\catcode `\_12\catcode `\%12\relax}%
\providecommand \@@startlink[1]{}%
\providecommand \@@endlink[0]{}%
\providecommand \url  [0]{\begingroup\@sanitize@url \@url }%
\providecommand \@url [1]{\endgroup\@href {#1}{\urlprefix }}%
\providecommand \urlprefix  [0]{URL }%
\providecommand \Eprint [0]{\href }%
\providecommand \doibase [0]{https://doi.org/}%
\providecommand \selectlanguage [0]{\@gobble}%
\providecommand \bibinfo  [0]{\@secondoftwo}%
\providecommand \bibfield  [0]{\@secondoftwo}%
\providecommand \translation [1]{[#1]}%
\providecommand \BibitemOpen [0]{}%
\providecommand \bibitemStop [0]{}%
\providecommand \bibitemNoStop [0]{.\EOS\space}%
\providecommand \EOS [0]{\spacefactor3000\relax}%
\providecommand \BibitemShut  [1]{\csname bibitem#1\endcsname}%
\let\auto@bib@innerbib\@empty
\bibitem [{\citenamefont {Atia}\ \emph {et~al.}(2021)\citenamefont {Atia}, \citenamefont {Fredberg}, \citenamefont {Gov},\ and\ \citenamefont {Pegoraro}}]{atiaAreCellJamming2021a}%
  \BibitemOpen
  \bibfield  {author} {\bibinfo {author} {\bibfnamefont {L.}~\bibnamefont {Atia}}, \bibinfo {author} {\bibfnamefont {J.~J.}\ \bibnamefont {Fredberg}}, \bibinfo {author} {\bibfnamefont {N.~S.}\ \bibnamefont {Gov}},\ and\ \bibinfo {author} {\bibfnamefont {A.~F.}\ \bibnamefont {Pegoraro}},\ }\bibfield  {title} {\bibinfo {title} {Are cell jamming and unjamming essential in tissue development?},\ }\href@noop {} {\bibfield  {journal} {\bibinfo  {journal} {Cells \& development}\ }\textbf {\bibinfo {volume} {168}},\ \bibinfo {pages} {203727} (\bibinfo {year} {2021})}\BibitemShut {NoStop}%
\bibitem [{\citenamefont {Lawson-Keister}\ and\ \citenamefont {Manning}(2021)}]{lawson2021jamming}%
  \BibitemOpen
  \bibfield  {author} {\bibinfo {author} {\bibfnamefont {E.}~\bibnamefont {Lawson-Keister}}\ and\ \bibinfo {author} {\bibfnamefont {M.~L.}\ \bibnamefont {Manning}},\ }\bibfield  {title} {\bibinfo {title} {Jamming and arrest of cell motion in biological tissues},\ }\href@noop {} {\bibfield  {journal} {\bibinfo  {journal} {Current Opinion in Cell Biology}\ }\textbf {\bibinfo {volume} {72}},\ \bibinfo {pages} {146} (\bibinfo {year} {2021})}\BibitemShut {NoStop}%
\bibitem [{\citenamefont {Cheung}\ and\ \citenamefont {Horne-Badovinac}(2025)}]{cheung2025collective}%
  \BibitemOpen
  \bibfield  {author} {\bibinfo {author} {\bibfnamefont {K.~J.}\ \bibnamefont {Cheung}}\ and\ \bibinfo {author} {\bibfnamefont {S.}~\bibnamefont {Horne-Badovinac}},\ }\bibfield  {title} {\bibinfo {title} {Collective migration modes in development, tissue repair and cancer},\ }\href@noop {} {\bibfield  {journal} {\bibinfo  {journal} {Nature Reviews Molecular Cell Biology}\ ,\ \bibinfo {pages} {1}} (\bibinfo {year} {2025})}\BibitemShut {NoStop}%
\bibitem [{\citenamefont {Lin}\ \emph {et~al.}(2025)\citenamefont {Lin}, \citenamefont {Yu},\ and\ \citenamefont {Pathak}}]{lin2025gradients}%
  \BibitemOpen
  \bibfield  {author} {\bibinfo {author} {\bibfnamefont {W.-J.}\ \bibnamefont {Lin}}, \bibinfo {author} {\bibfnamefont {H.}~\bibnamefont {Yu}},\ and\ \bibinfo {author} {\bibfnamefont {A.}~\bibnamefont {Pathak}},\ }\bibfield  {title} {\bibinfo {title} {Gradients in cell density and shape transitions drive collective cell migration into confining environments},\ }\href@noop {} {\bibfield  {journal} {\bibinfo  {journal} {Soft Matter}\ }\textbf {\bibinfo {volume} {21}},\ \bibinfo {pages} {719} (\bibinfo {year} {2025})}\BibitemShut {NoStop}%
\bibitem [{\citenamefont {Abercrombie}(1970)}]{abercrombie1970contact}%
  \BibitemOpen
  \bibfield  {author} {\bibinfo {author} {\bibfnamefont {M.}~\bibnamefont {Abercrombie}},\ }\bibfield  {title} {\bibinfo {title} {Contact inhibition in tissue culture},\ }\href@noop {} {\bibfield  {journal} {\bibinfo  {journal} {In vitro}\ }\textbf {\bibinfo {volume} {6}},\ \bibinfo {pages} {128} (\bibinfo {year} {1970})}\BibitemShut {NoStop}%
\bibitem [{\citenamefont {Theveneau}\ \emph {et~al.}(2010)\citenamefont {Theveneau}, \citenamefont {Marchant}, \citenamefont {Kuriyama}, \citenamefont {Gull}, \citenamefont {Moepps}, \citenamefont {Parsons},\ and\ \citenamefont {Mayor}}]{theveneau2010collective}%
  \BibitemOpen
  \bibfield  {author} {\bibinfo {author} {\bibfnamefont {E.}~\bibnamefont {Theveneau}}, \bibinfo {author} {\bibfnamefont {L.}~\bibnamefont {Marchant}}, \bibinfo {author} {\bibfnamefont {S.}~\bibnamefont {Kuriyama}}, \bibinfo {author} {\bibfnamefont {M.}~\bibnamefont {Gull}}, \bibinfo {author} {\bibfnamefont {B.}~\bibnamefont {Moepps}}, \bibinfo {author} {\bibfnamefont {M.}~\bibnamefont {Parsons}},\ and\ \bibinfo {author} {\bibfnamefont {R.}~\bibnamefont {Mayor}},\ }\bibfield  {title} {\bibinfo {title} {Collective chemotaxis requires contact-dependent cell polarity},\ }\href@noop {} {\bibfield  {journal} {\bibinfo  {journal} {Developmental cell}\ }\textbf {\bibinfo {volume} {19}},\ \bibinfo {pages} {39} (\bibinfo {year} {2010})}\BibitemShut {NoStop}%
\bibitem [{\citenamefont {Puliafito}\ \emph {et~al.}(2012)\citenamefont {Puliafito}, \citenamefont {Hufnagel}, \citenamefont {Neveu}, \citenamefont {Streichan}, \citenamefont {Sigal}, \citenamefont {Fygenson},\ and\ \citenamefont {Shraiman}}]{puliafito2012collective}%
  \BibitemOpen
  \bibfield  {author} {\bibinfo {author} {\bibfnamefont {A.}~\bibnamefont {Puliafito}}, \bibinfo {author} {\bibfnamefont {L.}~\bibnamefont {Hufnagel}}, \bibinfo {author} {\bibfnamefont {P.}~\bibnamefont {Neveu}}, \bibinfo {author} {\bibfnamefont {S.}~\bibnamefont {Streichan}}, \bibinfo {author} {\bibfnamefont {A.}~\bibnamefont {Sigal}}, \bibinfo {author} {\bibfnamefont {D.~K.}\ \bibnamefont {Fygenson}},\ and\ \bibinfo {author} {\bibfnamefont {B.~I.}\ \bibnamefont {Shraiman}},\ }\bibfield  {title} {\bibinfo {title} {Collective and single cell behavior in epithelial contact inhibition},\ }\href@noop {} {\bibfield  {journal} {\bibinfo  {journal} {Proceedings of the National Academy of Sciences}\ }\textbf {\bibinfo {volume} {109}},\ \bibinfo {pages} {739} (\bibinfo {year} {2012})}\BibitemShut {NoStop}%
\bibitem [{\citenamefont {Zimmermann}\ \emph {et~al.}(2016)\citenamefont {Zimmermann}, \citenamefont {Camley}, \citenamefont {Rappel},\ and\ \citenamefont {Levine}}]{zimmermann2016contact}%
  \BibitemOpen
  \bibfield  {author} {\bibinfo {author} {\bibfnamefont {J.}~\bibnamefont {Zimmermann}}, \bibinfo {author} {\bibfnamefont {B.~A.}\ \bibnamefont {Camley}}, \bibinfo {author} {\bibfnamefont {W.-J.}\ \bibnamefont {Rappel}},\ and\ \bibinfo {author} {\bibfnamefont {H.}~\bibnamefont {Levine}},\ }\bibfield  {title} {\bibinfo {title} {Contact inhibition of locomotion determines cell--cell and cell--substrate forces in tissues},\ }\href@noop {} {\bibfield  {journal} {\bibinfo  {journal} {Proceedings of the National Academy of Sciences}\ }\textbf {\bibinfo {volume} {113}},\ \bibinfo {pages} {2660} (\bibinfo {year} {2016})}\BibitemShut {NoStop}%
\bibitem [{\citenamefont {Li}\ \emph {et~al.}(2013)\citenamefont {Li}, \citenamefont {He}, \citenamefont {Zhao},\ and\ \citenamefont {Jiang}}]{li2013collective}%
  \BibitemOpen
  \bibfield  {author} {\bibinfo {author} {\bibfnamefont {L.}~\bibnamefont {Li}}, \bibinfo {author} {\bibfnamefont {Y.}~\bibnamefont {He}}, \bibinfo {author} {\bibfnamefont {M.}~\bibnamefont {Zhao}},\ and\ \bibinfo {author} {\bibfnamefont {J.}~\bibnamefont {Jiang}},\ }\bibfield  {title} {\bibinfo {title} {Collective cell migration: Implications for wound healing and cancer invasion},\ }\href@noop {} {\bibfield  {journal} {\bibinfo  {journal} {Burns \& trauma}\ }\textbf {\bibinfo {volume} {1}},\ \bibinfo {pages} {2321} (\bibinfo {year} {2013})}\BibitemShut {NoStop}%
\bibitem [{\citenamefont {Tian}\ \emph {et~al.}(2024)\citenamefont {Tian}, \citenamefont {Wang}, \citenamefont {Su}, \citenamefont {Huang}, \citenamefont {Zhang}, \citenamefont {Dou}, \citenamefont {Zhao}, \citenamefont {Cai}, \citenamefont {Pan}, \citenamefont {Bai} \emph {et~al.}}]{tian2024motility}%
  \BibitemOpen
  \bibfield  {author} {\bibinfo {author} {\bibfnamefont {C.}~\bibnamefont {Tian}}, \bibinfo {author} {\bibfnamefont {Y.}~\bibnamefont {Wang}}, \bibinfo {author} {\bibfnamefont {M.}~\bibnamefont {Su}}, \bibinfo {author} {\bibfnamefont {Y.}~\bibnamefont {Huang}}, \bibinfo {author} {\bibfnamefont {Y.}~\bibnamefont {Zhang}}, \bibinfo {author} {\bibfnamefont {J.}~\bibnamefont {Dou}}, \bibinfo {author} {\bibfnamefont {C.}~\bibnamefont {Zhao}}, \bibinfo {author} {\bibfnamefont {Y.}~\bibnamefont {Cai}}, \bibinfo {author} {\bibfnamefont {J.}~\bibnamefont {Pan}}, \bibinfo {author} {\bibfnamefont {S.}~\bibnamefont {Bai}}, \emph {et~al.},\ }\bibfield  {title} {\bibinfo {title} {Motility and tumor infiltration are key aspects of invariant natural killer t cell anti-tumor function},\ }\href@noop {} {\bibfield  {journal} {\bibinfo  {journal} {Nature communications}\ }\textbf {\bibinfo {volume} {15}},\ \bibinfo {pages} {1213} (\bibinfo {year} {2024})}\BibitemShut {NoStop}%
\bibitem [{\citenamefont {Park}\ \emph {et~al.}(2015)\citenamefont {Park}, \citenamefont {Kim}, \citenamefont {Bi}, \citenamefont {Mitchel}, \citenamefont {Qazvini}, \citenamefont {Tantisira}, \citenamefont {Park}, \citenamefont {McGill}, \citenamefont {Kim}, \citenamefont {Gweon} \emph {et~al.}}]{parkUnjammingCellShape2015a}%
  \BibitemOpen
  \bibfield  {author} {\bibinfo {author} {\bibfnamefont {J.-A.}\ \bibnamefont {Park}}, \bibinfo {author} {\bibfnamefont {J.~H.}\ \bibnamefont {Kim}}, \bibinfo {author} {\bibfnamefont {D.}~\bibnamefont {Bi}}, \bibinfo {author} {\bibfnamefont {J.~A.}\ \bibnamefont {Mitchel}}, \bibinfo {author} {\bibfnamefont {N.~T.}\ \bibnamefont {Qazvini}}, \bibinfo {author} {\bibfnamefont {K.}~\bibnamefont {Tantisira}}, \bibinfo {author} {\bibfnamefont {C.~Y.}\ \bibnamefont {Park}}, \bibinfo {author} {\bibfnamefont {M.}~\bibnamefont {McGill}}, \bibinfo {author} {\bibfnamefont {S.-H.}\ \bibnamefont {Kim}}, \bibinfo {author} {\bibfnamefont {B.}~\bibnamefont {Gweon}}, \emph {et~al.},\ }\bibfield  {title} {\bibinfo {title} {Unjamming and cell shape in the asthmatic airway epithelium},\ }\href@noop {} {\bibfield  {journal} {\bibinfo  {journal} {Nature materials}\ }\textbf {\bibinfo {volume} {14}},\ \bibinfo {pages} {1040} (\bibinfo {year} {2015})}\BibitemShut {NoStop}%
\bibitem [{\citenamefont {Xu}\ and\ \citenamefont {Wu}(2024)}]{xuSelfenhancedMobilityEnables2024}%
  \BibitemOpen
  \bibfield  {author} {\bibinfo {author} {\bibfnamefont {H.}~\bibnamefont {Xu}}\ and\ \bibinfo {author} {\bibfnamefont {Y.}~\bibnamefont {Wu}},\ }\bibfield  {title} {\bibinfo {title} {Self-enhanced mobility enables vortex pattern formation in living matter},\ }\href@noop {} {\bibfield  {journal} {\bibinfo  {journal} {Nature}\ }\textbf {\bibinfo {volume} {627}},\ \bibinfo {pages} {553} (\bibinfo {year} {2024})}\BibitemShut {NoStop}%
\bibitem [{\citenamefont {Chisolm}\ \emph {et~al.}(2025)\citenamefont {Chisolm}, \citenamefont {Guo}, \citenamefont {Subramaniam}, \citenamefont {Schulze},\ and\ \citenamefont {Angelini}}]{chisolmTransitionsCooperativeCrowdingdominated2025}%
  \BibitemOpen
  \bibfield  {author} {\bibinfo {author} {\bibfnamefont {S.~J.}\ \bibnamefont {Chisolm}}, \bibinfo {author} {\bibfnamefont {E.}~\bibnamefont {Guo}}, \bibinfo {author} {\bibfnamefont {V.}~\bibnamefont {Subramaniam}}, \bibinfo {author} {\bibfnamefont {K.~D.}\ \bibnamefont {Schulze}},\ and\ \bibinfo {author} {\bibfnamefont {T.~E.}\ \bibnamefont {Angelini}},\ }\bibfield  {title} {\bibinfo {title} {Transitions between cooperative and crowding-dominated collective motion in non-jammed mdck monolayers},\ }\href@noop {} {\bibfield  {journal} {\bibinfo  {journal} {Cells \& Development}\ }\textbf {\bibinfo {volume} {181}},\ \bibinfo {pages} {203989} (\bibinfo {year} {2025})}\BibitemShut {NoStop}%
\bibitem [{\citenamefont {Ruider}\ \emph {et~al.}(2024)\citenamefont {Ruider}, \citenamefont {Thijssen}, \citenamefont {Vannier}, \citenamefont {Paloschi}, \citenamefont {Sciortino}, \citenamefont {Doostmohammadi},\ and\ \citenamefont {Bausch}}]{ruider2024topological}%
  \BibitemOpen
  \bibfield  {author} {\bibinfo {author} {\bibfnamefont {I.}~\bibnamefont {Ruider}}, \bibinfo {author} {\bibfnamefont {K.}~\bibnamefont {Thijssen}}, \bibinfo {author} {\bibfnamefont {D.~R.}\ \bibnamefont {Vannier}}, \bibinfo {author} {\bibfnamefont {V.}~\bibnamefont {Paloschi}}, \bibinfo {author} {\bibfnamefont {A.}~\bibnamefont {Sciortino}}, \bibinfo {author} {\bibfnamefont {A.}~\bibnamefont {Doostmohammadi}},\ and\ \bibinfo {author} {\bibfnamefont {A.~R.}\ \bibnamefont {Bausch}},\ }\bibfield  {title} {\bibinfo {title} {Topological excitations govern ordering kinetics in endothelial cell layers},\ }\href@noop {} {\bibfield  {journal} {\bibinfo  {journal} {bioRxiv}\ ,\ \bibinfo {pages} {2024}} (\bibinfo {year} {2024})}\BibitemShut {NoStop}%
\bibitem [{\citenamefont {Serra-Picamal}\ \emph {et~al.}(2012)\citenamefont {Serra-Picamal}, \citenamefont {Conte}, \citenamefont {Vincent}, \citenamefont {Anon}, \citenamefont {Tambe}, \citenamefont {Bazellieres}, \citenamefont {Butler}, \citenamefont {Fredberg},\ and\ \citenamefont {Trepat}}]{serra2012mechanical}%
  \BibitemOpen
  \bibfield  {author} {\bibinfo {author} {\bibfnamefont {X.}~\bibnamefont {Serra-Picamal}}, \bibinfo {author} {\bibfnamefont {V.}~\bibnamefont {Conte}}, \bibinfo {author} {\bibfnamefont {R.}~\bibnamefont {Vincent}}, \bibinfo {author} {\bibfnamefont {E.}~\bibnamefont {Anon}}, \bibinfo {author} {\bibfnamefont {D.~T.}\ \bibnamefont {Tambe}}, \bibinfo {author} {\bibfnamefont {E.}~\bibnamefont {Bazellieres}}, \bibinfo {author} {\bibfnamefont {J.~P.}\ \bibnamefont {Butler}}, \bibinfo {author} {\bibfnamefont {J.~J.}\ \bibnamefont {Fredberg}},\ and\ \bibinfo {author} {\bibfnamefont {X.}~\bibnamefont {Trepat}},\ }\bibfield  {title} {\bibinfo {title} {Mechanical stress and cell density control cell proliferation in epithelial tissues},\ }\href@noop {} {\bibfield  {journal} {\bibinfo  {journal} {Nature Cell Biology}\ }\textbf {\bibinfo {volume} {14}},\ \bibinfo {pages} {540} (\bibinfo {year} {2012})}\BibitemShut {NoStop}%
\bibitem [{\citenamefont {Shahin-Shamsabadi}\ and\ \citenamefont {Selvaganapathy}(2020)}]{shahin2020tissue}%
  \BibitemOpen
  \bibfield  {author} {\bibinfo {author} {\bibfnamefont {A.}~\bibnamefont {Shahin-Shamsabadi}}\ and\ \bibinfo {author} {\bibfnamefont {P.}~\bibnamefont {Selvaganapathy}},\ }\bibfield  {title} {\bibinfo {title} {Tissue-in-a-tube: three-dimensional in vitro tissue constructs with integrated multimodal environmental stimulation},\ }\href@noop {} {\bibfield  {journal} {\bibinfo  {journal} {Materials Today Bio}\ }\textbf {\bibinfo {volume} {7}},\ \bibinfo {pages} {100070} (\bibinfo {year} {2020})}\BibitemShut {NoStop}%
\bibitem [{\citenamefont {Loza}\ \emph {et~al.}(2016)\citenamefont {Loza}, \citenamefont {Koride}, \citenamefont {Schimizzi}, \citenamefont {Li}, \citenamefont {Sun},\ and\ \citenamefont {Longmore}}]{loza2016cell}%
  \BibitemOpen
  \bibfield  {author} {\bibinfo {author} {\bibfnamefont {A.~J.}\ \bibnamefont {Loza}}, \bibinfo {author} {\bibfnamefont {S.}~\bibnamefont {Koride}}, \bibinfo {author} {\bibfnamefont {G.~V.}\ \bibnamefont {Schimizzi}}, \bibinfo {author} {\bibfnamefont {B.}~\bibnamefont {Li}}, \bibinfo {author} {\bibfnamefont {S.~X.}\ \bibnamefont {Sun}},\ and\ \bibinfo {author} {\bibfnamefont {G.~D.}\ \bibnamefont {Longmore}},\ }\bibfield  {title} {\bibinfo {title} {Cell density and actomyosin contractility control the organization of migrating collectives within an epithelium},\ }\href@noop {} {\bibfield  {journal} {\bibinfo  {journal} {Molecular Biology of the Cell}\ }\textbf {\bibinfo {volume} {27}},\ \bibinfo {pages} {3459} (\bibinfo {year} {2016})}\BibitemShut {NoStop}%
\bibitem [{\citenamefont {Thampi}\ \emph {et~al.}(2015)\citenamefont {Thampi}, \citenamefont {Doostmohammadi}, \citenamefont {Golestanian},\ and\ \citenamefont {Yeomans}}]{thampi2015intrinsic}%
  \BibitemOpen
  \bibfield  {author} {\bibinfo {author} {\bibfnamefont {S.~P.}\ \bibnamefont {Thampi}}, \bibinfo {author} {\bibfnamefont {A.}~\bibnamefont {Doostmohammadi}}, \bibinfo {author} {\bibfnamefont {R.}~\bibnamefont {Golestanian}},\ and\ \bibinfo {author} {\bibfnamefont {J.~M.}\ \bibnamefont {Yeomans}},\ }\bibfield  {title} {\bibinfo {title} {Intrinsic free energy in active nematics},\ }\href@noop {} {\bibfield  {journal} {\bibinfo  {journal} {Europhysics Letters}\ }\textbf {\bibinfo {volume} {112}},\ \bibinfo {pages} {28004} (\bibinfo {year} {2015})}\BibitemShut {NoStop}%
\bibitem [{\citenamefont {Doostmohammadi}\ \emph {et~al.}(2015)\citenamefont {Doostmohammadi}, \citenamefont {Thampi}, \citenamefont {Saw}, \citenamefont {Lim}, \citenamefont {Ladoux},\ and\ \citenamefont {Yeomans}}]{doostmohammadi2015celebrating}%
  \BibitemOpen
  \bibfield  {author} {\bibinfo {author} {\bibfnamefont {A.}~\bibnamefont {Doostmohammadi}}, \bibinfo {author} {\bibfnamefont {S.~P.}\ \bibnamefont {Thampi}}, \bibinfo {author} {\bibfnamefont {T.~B.}\ \bibnamefont {Saw}}, \bibinfo {author} {\bibfnamefont {C.~T.}\ \bibnamefont {Lim}}, \bibinfo {author} {\bibfnamefont {B.}~\bibnamefont {Ladoux}},\ and\ \bibinfo {author} {\bibfnamefont {J.~M.}\ \bibnamefont {Yeomans}},\ }\bibfield  {title} {\bibinfo {title} {Celebrating soft matter's 10th anniversary: Cell division: a source of active stress in cellular monolayers},\ }\href@noop {} {\bibfield  {journal} {\bibinfo  {journal} {Soft Matter}\ }\textbf {\bibinfo {volume} {11}},\ \bibinfo {pages} {7328} (\bibinfo {year} {2015})}\BibitemShut {NoStop}%
\bibitem [{\citenamefont {Fu}\ \emph {et~al.}(2012)\citenamefont {Fu}, \citenamefont {Tang}, \citenamefont {Liu}, \citenamefont {Huang}, \citenamefont {Hwa},\ and\ \citenamefont {Lenz}}]{fuStripeFormationBacterial2012}%
  \BibitemOpen
  \bibfield  {author} {\bibinfo {author} {\bibfnamefont {X.}~\bibnamefont {Fu}}, \bibinfo {author} {\bibfnamefont {L.-H.}\ \bibnamefont {Tang}}, \bibinfo {author} {\bibfnamefont {C.}~\bibnamefont {Liu}}, \bibinfo {author} {\bibfnamefont {J.-D.}\ \bibnamefont {Huang}}, \bibinfo {author} {\bibfnamefont {T.}~\bibnamefont {Hwa}},\ and\ \bibinfo {author} {\bibfnamefont {P.}~\bibnamefont {Lenz}},\ }\bibfield  {title} {\bibinfo {title} {Stripe formation in bacterial systems with density-suppressed motility},\ }\href@noop {} {\bibfield  {journal} {\bibinfo  {journal} {Physical review letters}\ }\textbf {\bibinfo {volume} {108}},\ \bibinfo {pages} {198102} (\bibinfo {year} {2012})}\BibitemShut {NoStop}%
\bibitem [{\citenamefont {Cates}\ and\ \citenamefont {Tailleur}(2015)}]{cates2015motility}%
  \BibitemOpen
  \bibfield  {author} {\bibinfo {author} {\bibfnamefont {M.~E.}\ \bibnamefont {Cates}}\ and\ \bibinfo {author} {\bibfnamefont {J.}~\bibnamefont {Tailleur}},\ }\bibfield  {title} {\bibinfo {title} {Motility-induced phase separation},\ }\href@noop {} {\bibfield  {journal} {\bibinfo  {journal} {Annual Review of Condensed Matter Physics}\ }\textbf {\bibinfo {volume} {6}},\ \bibinfo {pages} {219} (\bibinfo {year} {2015})}\BibitemShut {NoStop}%
\bibitem [{\citenamefont {Cates}\ \emph {et~al.}(2010)\citenamefont {Cates}, \citenamefont {Marenduzzo}, \citenamefont {Pagonabarraga},\ and\ \citenamefont {Tailleur}}]{catesArrestedPhaseSeparation2010}%
  \BibitemOpen
  \bibfield  {author} {\bibinfo {author} {\bibfnamefont {M.~E.}\ \bibnamefont {Cates}}, \bibinfo {author} {\bibfnamefont {D.}~\bibnamefont {Marenduzzo}}, \bibinfo {author} {\bibfnamefont {I.}~\bibnamefont {Pagonabarraga}},\ and\ \bibinfo {author} {\bibfnamefont {J.}~\bibnamefont {Tailleur}},\ }\bibfield  {title} {\bibinfo {title} {Arrested phase separation in reproducing bacteria creates a generic route to pattern formation},\ }\href@noop {} {\bibfield  {journal} {\bibinfo  {journal} {Proceedings of the National Academy of Sciences}\ }\textbf {\bibinfo {volume} {107}},\ \bibinfo {pages} {11715} (\bibinfo {year} {2010})}\BibitemShut {NoStop}%
\bibitem [{\citenamefont {Marchetti}\ \emph {et~al.}(2013)\citenamefont {Marchetti}, \citenamefont {Joanny}, \citenamefont {Ramaswamy}, \citenamefont {Liverpool}, \citenamefont {Prost}, \citenamefont {Rao},\ and\ \citenamefont {Simha}}]{marchetti2013hydrodynamics}%
  \BibitemOpen
  \bibfield  {author} {\bibinfo {author} {\bibfnamefont {M.~C.}\ \bibnamefont {Marchetti}}, \bibinfo {author} {\bibfnamefont {J.-F.}\ \bibnamefont {Joanny}}, \bibinfo {author} {\bibfnamefont {S.}~\bibnamefont {Ramaswamy}}, \bibinfo {author} {\bibfnamefont {T.~B.}\ \bibnamefont {Liverpool}}, \bibinfo {author} {\bibfnamefont {J.}~\bibnamefont {Prost}}, \bibinfo {author} {\bibfnamefont {M.}~\bibnamefont {Rao}},\ and\ \bibinfo {author} {\bibfnamefont {R.~A.}\ \bibnamefont {Simha}},\ }\bibfield  {title} {\bibinfo {title} {Hydrodynamics of soft active matter},\ }\href@noop {} {\bibfield  {journal} {\bibinfo  {journal} {Reviews of Modern Physics}\ }\textbf {\bibinfo {volume} {85}},\ \bibinfo {pages} {1143} (\bibinfo {year} {2013})}\BibitemShut {NoStop}%
\bibitem [{\citenamefont {Roycroft}\ and\ \citenamefont {Mayor}(2016)}]{roycroftMolecularBasisContact2016}%
  \BibitemOpen
  \bibfield  {author} {\bibinfo {author} {\bibfnamefont {A.}~\bibnamefont {Roycroft}}\ and\ \bibinfo {author} {\bibfnamefont {R.}~\bibnamefont {Mayor}},\ }\bibfield  {title} {\bibinfo {title} {Molecular basis of contact inhibition of locomotion},\ }\href@noop {} {\bibfield  {journal} {\bibinfo  {journal} {Cellular and Molecular Life Sciences}\ }\textbf {\bibinfo {volume} {73}},\ \bibinfo {pages} {1119} (\bibinfo {year} {2016})}\BibitemShut {NoStop}%
\bibitem [{\citenamefont {Mueller}\ \emph {et~al.}(2019)\citenamefont {Mueller}, \citenamefont {Yeomans},\ and\ \citenamefont {Doostmohammadi}}]{PhysRevLett.122.048004}%
  \BibitemOpen
  \bibfield  {author} {\bibinfo {author} {\bibfnamefont {R.}~\bibnamefont {Mueller}}, \bibinfo {author} {\bibfnamefont {J.~M.}\ \bibnamefont {Yeomans}},\ and\ \bibinfo {author} {\bibfnamefont {A.}~\bibnamefont {Doostmohammadi}},\ }\bibfield  {title} {\bibinfo {title} {Emergence of active nematic behavior in monolayers of isotropic cells},\ }\href {https://doi.org/10.1103/PhysRevLett.122.048004} {\bibfield  {journal} {\bibinfo  {journal} {Phys. Rev. Lett.}\ }\textbf {\bibinfo {volume} {122}},\ \bibinfo {pages} {048004} (\bibinfo {year} {2019})}\BibitemShut {NoStop}%
\bibitem [{\citenamefont {Angelini}\ \emph {et~al.}(2011)\citenamefont {Angelini}, \citenamefont {Hannezo}, \citenamefont {Trepat}, \citenamefont {Marquez}, \citenamefont {Fredberg},\ and\ \citenamefont {Weitz}}]{angelini2011glass}%
  \BibitemOpen
  \bibfield  {author} {\bibinfo {author} {\bibfnamefont {T.~E.}\ \bibnamefont {Angelini}}, \bibinfo {author} {\bibfnamefont {E.}~\bibnamefont {Hannezo}}, \bibinfo {author} {\bibfnamefont {X.}~\bibnamefont {Trepat}}, \bibinfo {author} {\bibfnamefont {M.}~\bibnamefont {Marquez}}, \bibinfo {author} {\bibfnamefont {J.~J.}\ \bibnamefont {Fredberg}},\ and\ \bibinfo {author} {\bibfnamefont {D.~A.}\ \bibnamefont {Weitz}},\ }\bibfield  {title} {\bibinfo {title} {Glass-like dynamics of collective cell migration},\ }\href@noop {} {\bibfield  {journal} {\bibinfo  {journal} {Proceedings of the National Academy of Sciences}\ }\textbf {\bibinfo {volume} {108}},\ \bibinfo {pages} {4714} (\bibinfo {year} {2011})}\BibitemShut {NoStop}%
\bibitem [{\citenamefont {Atia}\ \emph {et~al.}(2018)\citenamefont {Atia}, \citenamefont {Bi}, \citenamefont {Sharma}, \citenamefont {Mitchel}, \citenamefont {Gweon}, \citenamefont {A.~Koehler}, \citenamefont {DeCamp}, \citenamefont {Lan}, \citenamefont {Kim}, \citenamefont {Hirsch} \emph {et~al.}}]{atia2018geometric}%
  \BibitemOpen
  \bibfield  {author} {\bibinfo {author} {\bibfnamefont {L.}~\bibnamefont {Atia}}, \bibinfo {author} {\bibfnamefont {D.}~\bibnamefont {Bi}}, \bibinfo {author} {\bibfnamefont {Y.}~\bibnamefont {Sharma}}, \bibinfo {author} {\bibfnamefont {J.~A.}\ \bibnamefont {Mitchel}}, \bibinfo {author} {\bibfnamefont {B.}~\bibnamefont {Gweon}}, \bibinfo {author} {\bibfnamefont {S.}~\bibnamefont {A.~Koehler}}, \bibinfo {author} {\bibfnamefont {S.~J.}\ \bibnamefont {DeCamp}}, \bibinfo {author} {\bibfnamefont {B.}~\bibnamefont {Lan}}, \bibinfo {author} {\bibfnamefont {J.~H.}\ \bibnamefont {Kim}}, \bibinfo {author} {\bibfnamefont {R.}~\bibnamefont {Hirsch}}, \emph {et~al.},\ }\bibfield  {title} {\bibinfo {title} {Geometric constraints during epithelial jamming},\ }\href@noop {} {\bibfield  {journal} {\bibinfo  {journal} {Nature physics}\ }\textbf {\bibinfo {volume} {14}},\ \bibinfo {pages} {613} (\bibinfo {year} {2018})}\BibitemShut {NoStop}%
\bibitem [{\citenamefont {Thielicke}(2014)}]{thielickePIVlabUserfriendlyAffordable2014}%
  \BibitemOpen
  \bibfield  {author} {\bibinfo {author} {\bibfnamefont {W.}~\bibnamefont {Thielicke}},\ }\bibfield  {title} {\bibinfo {title} {Pivlab –towards user friendly, affordable and accurate digital particle image velocimetry in matlab},\ }\href {https://doi.org/http://doi.org/10.5334/jors.bl} {\bibfield  {journal} {\bibinfo  {journal} {Journal of Open Research Software}\ }\textbf {\bibinfo {volume} {2}},\ \bibinfo {pages} {30} (\bibinfo {year} {2014})}\BibitemShut {NoStop}%
\bibitem [{\citenamefont {Nier}\ \emph {et~al.}(2016{\natexlab{a}})\citenamefont {Nier}, \citenamefont {Jain}, \citenamefont {Lim}, \citenamefont {Ishihara}, \citenamefont {Ladoux},\ and\ \citenamefont {Marcq}}]{nier2016inference}%
  \BibitemOpen
  \bibfield  {author} {\bibinfo {author} {\bibfnamefont {V.}~\bibnamefont {Nier}}, \bibinfo {author} {\bibfnamefont {S.}~\bibnamefont {Jain}}, \bibinfo {author} {\bibfnamefont {C.~T.}\ \bibnamefont {Lim}}, \bibinfo {author} {\bibfnamefont {S.}~\bibnamefont {Ishihara}}, \bibinfo {author} {\bibfnamefont {B.}~\bibnamefont {Ladoux}},\ and\ \bibinfo {author} {\bibfnamefont {P.}~\bibnamefont {Marcq}},\ }\bibfield  {title} {\bibinfo {title} {Inference of internal stress in a cell monolayer},\ }\href@noop {} {\bibfield  {journal} {\bibinfo  {journal} {Biophysical journal}\ }\textbf {\bibinfo {volume} {110}},\ \bibinfo {pages} {1625} (\bibinfo {year} {2016}{\natexlab{a}})}\BibitemShut {NoStop}%
\bibitem [{\citenamefont {Li}(2008)}]{li2008image}%
  \BibitemOpen
  \bibfield  {author} {\bibinfo {author} {\bibfnamefont {K.}~\bibnamefont {Li}},\ }\href@noop {} {\bibinfo {title} {The image stabilizer plugin for imagej}},\ \bibinfo {howpublished} {\url{https://www.cs.cmu.edu/~kangli/code/Image_Stabilizer.html}} (\bibinfo {year} {2008})\BibitemShut {NoStop}%
\bibitem [{\citenamefont {Nier}\ \emph {et~al.}(2016{\natexlab{b}})\citenamefont {Nier}, \citenamefont {Jain}, \citenamefont {Lim}, \citenamefont {Ishihara}, \citenamefont {Ladoux},\ and\ \citenamefont {Marcq}}]{NIER20161625}%
  \BibitemOpen
  \bibfield  {author} {\bibinfo {author} {\bibfnamefont {V.}~\bibnamefont {Nier}}, \bibinfo {author} {\bibfnamefont {S.}~\bibnamefont {Jain}}, \bibinfo {author} {\bibfnamefont {C.~T.}\ \bibnamefont {Lim}}, \bibinfo {author} {\bibfnamefont {S.}~\bibnamefont {Ishihara}}, \bibinfo {author} {\bibfnamefont {B.}~\bibnamefont {Ladoux}},\ and\ \bibinfo {author} {\bibfnamefont {P.}~\bibnamefont {Marcq}},\ }\bibfield  {title} {\bibinfo {title} {Inference of internal stress in a cell monolayer},\ }\href@noop {} {\bibfield  {journal} {\bibinfo  {journal} {Biophysical journal}\ }\textbf {\bibinfo {volume} {110}},\ \bibinfo {pages} {1625} (\bibinfo {year} {2016}{\natexlab{b}})}\BibitemShut {NoStop}%
\bibitem [{\citenamefont {Mueller}\ and\ \citenamefont {Doostmohammadi}(2021)}]{muellerPhaseFieldModels2021}%
  \BibitemOpen
  \bibfield  {author} {\bibinfo {author} {\bibfnamefont {R.}~\bibnamefont {Mueller}}\ and\ \bibinfo {author} {\bibfnamefont {A.}~\bibnamefont {Doostmohammadi}},\ }\bibfield  {title} {\bibinfo {title} {Phase field models of active matter},\ }\href@noop {} {\bibfield  {journal} {\bibinfo  {journal} {arXiv preprint arXiv:2102.05557}\ } (\bibinfo {year} {2021})}\BibitemShut {NoStop}%
\bibitem [{\citenamefont {Monfared}\ \emph {et~al.}(2025)\citenamefont {Monfared}, \citenamefont {Arda{\v{s}}eva},\ and\ \citenamefont {Doostmohammadi}}]{monfared2025multi}%
  \BibitemOpen
  \bibfield  {author} {\bibinfo {author} {\bibfnamefont {S.}~\bibnamefont {Monfared}}, \bibinfo {author} {\bibfnamefont {A.}~\bibnamefont {Arda{\v{s}}eva}},\ and\ \bibinfo {author} {\bibfnamefont {A.}~\bibnamefont {Doostmohammadi}},\ }\bibfield  {title} {\bibinfo {title} {Multi-phase-field models of biological tissues},\ }\href@noop {} {\bibfield  {journal} {\bibinfo  {journal} {arXiv preprint arXiv:2503.05053}\ } (\bibinfo {year} {2025})}\BibitemShut {NoStop}%
\bibitem [{\citenamefont {Chiang}\ \emph {et~al.}(2024)\citenamefont {Chiang}, \citenamefont {Hopkins}, \citenamefont {Loewe}, \citenamefont {Marenduzzo},\ and\ \citenamefont {Marchetti}}]{PhysRevE.110.044403}%
  \BibitemOpen
  \bibfield  {author} {\bibinfo {author} {\bibfnamefont {M.}~\bibnamefont {Chiang}}, \bibinfo {author} {\bibfnamefont {A.}~\bibnamefont {Hopkins}}, \bibinfo {author} {\bibfnamefont {B.}~\bibnamefont {Loewe}}, \bibinfo {author} {\bibfnamefont {D.}~\bibnamefont {Marenduzzo}},\ and\ \bibinfo {author} {\bibfnamefont {M.~C.}\ \bibnamefont {Marchetti}},\ }\bibfield  {title} {\bibinfo {title} {Multiphase field model of cells on a substrate: From three dimensional to two dimensional},\ }\href {https://doi.org/10.1103/PhysRevE.110.044403} {\bibfield  {journal} {\bibinfo  {journal} {Phys. Rev. E}\ }\textbf {\bibinfo {volume} {110}},\ \bibinfo {pages} {044403} (\bibinfo {year} {2024})}\BibitemShut {NoStop}%
\bibitem [{\citenamefont {Palmieri}\ \emph {et~al.}(2015)\citenamefont {Palmieri}, \citenamefont {Bresler}, \citenamefont {Wirtz},\ and\ \citenamefont {Grant}}]{palmieriMultipleScaleModel2015}%
  \BibitemOpen
  \bibfield  {author} {\bibinfo {author} {\bibfnamefont {B.}~\bibnamefont {Palmieri}}, \bibinfo {author} {\bibfnamefont {Y.}~\bibnamefont {Bresler}}, \bibinfo {author} {\bibfnamefont {D.}~\bibnamefont {Wirtz}},\ and\ \bibinfo {author} {\bibfnamefont {M.}~\bibnamefont {Grant}},\ }\bibfield  {title} {\bibinfo {title} {Multiple scale model for cell migration in monolayers: elastic mismatch between cells enhances motility},\ }\href@noop {} {\bibfield  {journal} {\bibinfo  {journal} {Scientific Reports}\ }\textbf {\bibinfo {volume} {5}},\ \bibinfo {pages} {11745} (\bibinfo {year} {2015})}\BibitemShut {NoStop}%
\bibitem [{\citenamefont {Peyret}\ \emph {et~al.}(2019)\citenamefont {Peyret}, \citenamefont {Mueller}, \citenamefont {d’Alessandro}, \citenamefont {Begnaud}, \citenamefont {Marcq}, \citenamefont {M{\`e}ge}, \citenamefont {Yeomans}, \citenamefont {Doostmohammadi},\ and\ \citenamefont {Ladoux}}]{peyretSustainedOscillationsEpithelial2019}%
  \BibitemOpen
  \bibfield  {author} {\bibinfo {author} {\bibfnamefont {G.}~\bibnamefont {Peyret}}, \bibinfo {author} {\bibfnamefont {R.}~\bibnamefont {Mueller}}, \bibinfo {author} {\bibfnamefont {J.}~\bibnamefont {d’Alessandro}}, \bibinfo {author} {\bibfnamefont {S.}~\bibnamefont {Begnaud}}, \bibinfo {author} {\bibfnamefont {P.}~\bibnamefont {Marcq}}, \bibinfo {author} {\bibfnamefont {R.-M.}\ \bibnamefont {M{\`e}ge}}, \bibinfo {author} {\bibfnamefont {J.~M.}\ \bibnamefont {Yeomans}}, \bibinfo {author} {\bibfnamefont {A.}~\bibnamefont {Doostmohammadi}},\ and\ \bibinfo {author} {\bibfnamefont {B.}~\bibnamefont {Ladoux}},\ }\bibfield  {title} {\bibinfo {title} {Sustained oscillations of epithelial cell sheets},\ }\href@noop {} {\bibfield  {journal} {\bibinfo  {journal} {Biophysical Journal}\ }\textbf {\bibinfo {volume} {117}},\ \bibinfo {pages} {464} (\bibinfo {year} {2019})}\BibitemShut {NoStop}%
\bibitem [{\citenamefont {Cates}\ and\ \citenamefont {Tjhung}(2018)}]{Cates_Tjhung_2018}%
  \BibitemOpen
  \bibfield  {author} {\bibinfo {author} {\bibfnamefont {M.~E.}\ \bibnamefont {Cates}}\ and\ \bibinfo {author} {\bibfnamefont {E.}~\bibnamefont {Tjhung}},\ }\bibfield  {title} {\bibinfo {title} {Theories of binary fluid mixtures: from phase-separation kinetics to active emulsions},\ }\href {https://doi.org/10.1017/jfm.2017.832} {\bibfield  {journal} {\bibinfo  {journal} {Journal of Fluid Mechanics}\ }\textbf {\bibinfo {volume} {836}},\ \bibinfo {pages} {P1} (\bibinfo {year} {2018})}\BibitemShut {NoStop}%
\bibitem [{\citenamefont {Garcia}\ \emph {et~al.}(2015)\citenamefont {Garcia}, \citenamefont {Hannezo}, \citenamefont {Elgeti}, \citenamefont {Joanny}, \citenamefont {Silberzan},\ and\ \citenamefont {Gov}}]{garciaPhysicsActiveJamming2015}%
  \BibitemOpen
  \bibfield  {author} {\bibinfo {author} {\bibfnamefont {S.}~\bibnamefont {Garcia}}, \bibinfo {author} {\bibfnamefont {E.}~\bibnamefont {Hannezo}}, \bibinfo {author} {\bibfnamefont {J.}~\bibnamefont {Elgeti}}, \bibinfo {author} {\bibfnamefont {J.-F.}\ \bibnamefont {Joanny}}, \bibinfo {author} {\bibfnamefont {P.}~\bibnamefont {Silberzan}},\ and\ \bibinfo {author} {\bibfnamefont {N.~S.}\ \bibnamefont {Gov}},\ }\bibfield  {title} {\bibinfo {title} {Physics of active jamming during collective cellular motion in a monolayer},\ }\href@noop {} {\bibfield  {journal} {\bibinfo  {journal} {Proceedings of the National Academy of Sciences}\ }\textbf {\bibinfo {volume} {112}},\ \bibinfo {pages} {15314} (\bibinfo {year} {2015})}\BibitemShut {NoStop}%
\bibitem [{\citenamefont {Zehnder}\ \emph {et~al.}(2015)\citenamefont {Zehnder}, \citenamefont {Wiatt}, \citenamefont {Uruena}, \citenamefont {Dunn}, \citenamefont {Sawyer},\ and\ \citenamefont {Angelini}}]{zehnderMulticellularDensityFluctuations2015}%
  \BibitemOpen
  \bibfield  {author} {\bibinfo {author} {\bibfnamefont {S.~M.}\ \bibnamefont {Zehnder}}, \bibinfo {author} {\bibfnamefont {M.~K.}\ \bibnamefont {Wiatt}}, \bibinfo {author} {\bibfnamefont {J.~M.}\ \bibnamefont {Uruena}}, \bibinfo {author} {\bibfnamefont {A.~C.}\ \bibnamefont {Dunn}}, \bibinfo {author} {\bibfnamefont {W.~G.}\ \bibnamefont {Sawyer}},\ and\ \bibinfo {author} {\bibfnamefont {T.~E.}\ \bibnamefont {Angelini}},\ }\bibfield  {title} {\bibinfo {title} {Multicellular density fluctuations in epithelial monolayers},\ }\href@noop {} {\bibfield  {journal} {\bibinfo  {journal} {Physical Review E}\ }\textbf {\bibinfo {volume} {92}},\ \bibinfo {pages} {032729} (\bibinfo {year} {2015})}\BibitemShut {NoStop}%
\bibitem [{\citenamefont {Vishwakarma}\ and\ \citenamefont {Di~Russo}(2019)}]{vishwakarmaWhyDoesEpithelia2019}%
  \BibitemOpen
  \bibfield  {author} {\bibinfo {author} {\bibfnamefont {M.}~\bibnamefont {Vishwakarma}}\ and\ \bibinfo {author} {\bibfnamefont {J.}~\bibnamefont {Di~Russo}},\ }\bibfield  {title} {\bibinfo {title} {Why does epithelia display heterogeneity? bridging physical and biological concepts},\ }\href@noop {} {\bibfield  {journal} {\bibinfo  {journal} {Biophysical Reviews}\ }\textbf {\bibinfo {volume} {11}},\ \bibinfo {pages} {683} (\bibinfo {year} {2019})}\BibitemShut {NoStop}%
\bibitem [{\citenamefont {Tambe}\ \emph {et~al.}(2011)\citenamefont {Tambe}, \citenamefont {Hardin}, \citenamefont {Angelini}, \citenamefont {Rajendran}, \citenamefont {Park}, \citenamefont {Serra-Picamal}, \citenamefont {Zhou}, \citenamefont {Zaman}, \citenamefont {Butler}, \citenamefont {Weitz}, \citenamefont {Fredberg},\ and\ \citenamefont {Trepat}}]{tambeCollectiveCellGuidance2011}%
  \BibitemOpen
  \bibfield  {author} {\bibinfo {author} {\bibfnamefont {D.~T.}\ \bibnamefont {Tambe}}, \bibinfo {author} {\bibfnamefont {C.~C.}\ \bibnamefont {Hardin}}, \bibinfo {author} {\bibfnamefont {T.~E.}\ \bibnamefont {Angelini}}, \bibinfo {author} {\bibfnamefont {K.}~\bibnamefont {Rajendran}}, \bibinfo {author} {\bibfnamefont {C.~Y.}\ \bibnamefont {Park}}, \bibinfo {author} {\bibfnamefont {X.}~\bibnamefont {Serra-Picamal}}, \bibinfo {author} {\bibfnamefont {E.~H.}\ \bibnamefont {Zhou}}, \bibinfo {author} {\bibfnamefont {M.~H.}\ \bibnamefont {Zaman}}, \bibinfo {author} {\bibfnamefont {J.~P.}\ \bibnamefont {Butler}}, \bibinfo {author} {\bibfnamefont {D.~A.}\ \bibnamefont {Weitz}}, \bibinfo {author} {\bibfnamefont {J.~J.}\ \bibnamefont {Fredberg}},\ and\ \bibinfo {author} {\bibfnamefont {X.}~\bibnamefont {Trepat}},\ }\bibfield  {title} {\bibinfo {title} {Collective cell guidance by cooperative intercellular forces},\ }\href@noop {} {\bibfield  {journal} {\bibinfo  {journal} {Nature Materials}\ }\textbf {\bibinfo
  {volume} {10}},\ \bibinfo {pages} {469} (\bibinfo {year} {2011})}\BibitemShut {NoStop}%
\bibitem [{\citenamefont {Bra{\'n}ka}\ and\ \citenamefont {Heyes}(2011)}]{brankaPairCorrelationFunction2011}%
  \BibitemOpen
  \bibfield  {author} {\bibinfo {author} {\bibfnamefont {A.~C.}\ \bibnamefont {Bra{\'n}ka}}\ and\ \bibinfo {author} {\bibfnamefont {D.~M.}\ \bibnamefont {Heyes}},\ }\bibfield  {title} {\bibinfo {title} {Pair correlation function of soft-sphere fluids},\ }\href@noop {} {\bibfield  {journal} {\bibinfo  {journal} {The Journal of Chemical Physics}\ }\textbf {\bibinfo {volume} {134}},\ \bibinfo {pages} {064115} (\bibinfo {year} {2011})}\BibitemShut {NoStop}%
\bibitem [{\citenamefont {Cates}\ and\ \citenamefont {Tailleur}(2013)}]{catesWhenAreActive2013}%
  \BibitemOpen
  \bibfield  {author} {\bibinfo {author} {\bibfnamefont {M.~E.}\ \bibnamefont {Cates}}\ and\ \bibinfo {author} {\bibfnamefont {J.}~\bibnamefont {Tailleur}},\ }\bibfield  {title} {\bibinfo {title} {When are active brownian particles and run-and-tumble particles equivalent? consequences for motility-induced phase separation},\ }\href@noop {} {\bibfield  {journal} {\bibinfo  {journal} {EPL (Europhysics Letters)}\ }\textbf {\bibinfo {volume} {101}},\ \bibinfo {pages} {20010} (\bibinfo {year} {2013})}\BibitemShut {NoStop}%
\end{thebibliography}%

\clearpage
\appendix

\section{Implementation of multi-phase field model}
We numerically evolve the phase field model using a finite-difference method. Phase fields are evolved on a uniform square grid with periodic boundaries. Spatial derivatives use second-order central differences, and the Laplacian uses the standard five-point stencil. Free-energy terms in Eq.~\eqref{eq:freeenergy} are integrated by Riemann sums over the grid. Time integration is explicit (forward Euler), using the update rule $\phi_i(t+\Delta t) = \phi_i(t) + \Delta t\, \partial_t \phi_i$.

Initial conditions are disks of radius $R$ with random positions. The parameter setup is included in table \ref{tab:sim_params} and table \ref{tab:sim_sweeps}. $N$ and $L$ set the packing fraction. We enforce periodicity by wrap-around indexing and choose $\Delta t$ to satisfy stability for the chosen grid.



\section{Phase-field simulation parameters and parameter variations}
\begin{table}[ht]
\centering
\begin{minipage}{1\linewidth}
\caption{Phase-field simulation parameters used in Fig.~\ref{fig:4}}
\label{tab:sim_params}
\begin{ruledtabular}
\begin{tabular}{lll}
Symbol & Description & Value \\
\hline
$R$ & relax cell radius & $8$ \\
$\lambda$ & Interface width & $3$ \\
$\gamma$ & Interfacial energy & $0.07$ \\
$\mu$ & Area-penalty strength & $5$ \\
$\kappa$ & Steric repulsion strength & $0.3$ \\
$\omega$ & Cell--cell adhesion strength & $1\times10^{-3}$ \\
$\zeta$ & Activity (extensile) & $1.2\times10^{-2}$ \\
$\xi$ & Substrate friction & $1.0$ \\
$\Delta x$ & Grid spacing & $1$ \\
$\Delta t$ & Time step & $2\times 10^{-3}\,\tau$ \\
$L$ & System size (side) & $2400$ \\
$N$ & Number of cells & $40107$ \\
BC & Boundary condition & periodic \\
\end{tabular}
\end{ruledtabular}
\end{minipage}

\vspace{3ex}

\begin{minipage}{1\linewidth}
\caption{Parameter variations used across figures. Only the listed parameter differs from the base case in Table~\ref{tab:sim_params}.}
\label{tab:sim_sweeps}
\begin{ruledtabular}
\begin{tabular}{llll}
Figure & Varied parameter & Values & Purpose \\
\hline
Fig.~\ref{fig:5}A & $\zeta$ & $1.2\times10^{-2}$ & Reduce activity \\
Fig.~\ref{fig:5}C & Packing fraction $\phi$ & $1.0$ & Weaken crowding \\
Fig.~\ref{fig:4}C & System size $L$ & $500,\,1000,\,2400$ & Probe \(S(q)\) \\
\end{tabular}
\end{ruledtabular}
\end{minipage}
\end{table}

\end{document}